\documentclass[a4paper,11pt]{article}

\usepackage{jinstpub} 
\usepackage{subcaption}
\usepackage{lineno}
\usepackage{amsmath}
\usepackage{multirow}
\usepackage{upgreek}
\usepackage{makecell}
\usepackage{booktabs}
\usepackage{threeparttable}
\title{\boldmath Simulation of a Xe-based X-ray Polarimeter at 10-30~keV}

\author[a]{J.~Zhang,}
\author[a]{X.~Cai,}
\author[a,1]{Y.~Huang,\note{Corresponding author.}}
\author[b]{Q.~Liu,}
\author[a]{F.~Xie,}
\author[c]{J.~Li}
\author[a,2]{and H.~Liu,\note{Corresponding author.}}

\affiliation[a]{Guangxi Key Laboratory for Relativistic Astrophysics, School of Physical Science and Technology, Guangxi University, Nanning 530004, China}
\affiliation[b]{School of Physical Science, University of Chinese Academy of Sciences, Beijing 100049, China}
\affiliation[c]{Key Laboratory of Particle Astrophysics, Institute of High Energy Physics, Chinese Academy of Sciences, Beijing 100049, China}

\emailAdd{huangyb@gxu.edu.cn}
\emailAdd{liuhb@gxu.edu.cn}

\abstract{Polarization detection of X-rays is a non-negligible topic to astrophysical observation. Many polarization detection methods have been well developed for X-rays in the energy range below 10~keV, while the detection at 10-30~keV is rarely discussed. This paper presents a simulation study of a Xe-based gas pixel detector, which can achieve the polarization detection of X-rays at 10-30~keV. To verify the emission angle distribution of photoelectrons, different electromagnetic models in Geant4 were investigated. After a necessary modification by considering the missing factor when sampling the emission angle, a good agreement can be achieved. Moreover, the detection capability of 20~keV polarized photons was discussed and the modulation factor could be 43\%. }

\keywords{X-ray detectors, Polarimeters, Detector modelling and simulations I (interaction of radiation with matter, interaction of photons with matter, interaction of hadrons with matter, etc), Simulation methods and programs}

\begin{document}
\maketitle
\flushbottom

\section{Introduction}
\label{sec:intro}

Multidimensional observation is necessary for a comprehensive understanding of the universe. For instance, the sensitive X-ray polarization detection may provide additional evidence to explain the structure, as well as the physical processes of black holes, neutron stars, and all types of X-ray sources in the universe~\cite{X-rayScience,Bhattacharya1991,Cornelisse2003}. Therefore, it is apparent that an enhanced amount of information on the polarization of X-rays will be required in the future.

In the past few decades, the polarization detection of X-rays is almost a blank field in astronomical observation. In 1975, the polarization measurements of the Crab Nebula which were performed by the Orbiting Solar Observatory (OSO-8) satellite~\cite{Weisskopf1976} demonstrated the origin of the Crab Nebula’s X-ray emission synchrotron radiation. Decades after that, a CubeSat named PolarLight was launched into a sun-synchronous orbit on October 29, 2018, carrying a Gas Pixel Detector (GPD) with a collimator~\cite{PolarLight}. PolarLight focuses on the X-ray polarization detection of relatively strongly bright sources such as the Crab Nebula~\cite{Feng2020}. The re-opening of the polarization detection window in X-rays represents a renewal of the scientific interest to acquire significant insights from the polarization detection measurements of X-ray sources in the universe. As a result, not only the launched space observation missions such as IXPE~\cite{IXPE}, but also various X-ray polarization detection missions such as PRAXyS~\cite{PRAXyS}, eXTP~\cite{eXTP} and POLAR-2~\cite{POLAR-2} are planned in the near future. 

Low Energy Polarization Detector (LPD) is a GPD for X-ray polarization detection in POLAR-2, which has a large field of view and large area. LPD mainly focuses on the energy range from 2~keV to 30~keV. The polarization detection of X-rays in the energy range below 10~keV has been well studied, while the polarization detection at 10-30~keV is rarely discussed in the previous. To cover the wide energy range, two different working gases will be applied, respectively. The purpose of this paper is to study the polarization detection capability of X-rays at 10-30~keV using Xe as the main working gas. This study would focus on the investigation of the detection feasibility in simulation. As for the further experimental research, which was planned to carry out in the future, it is not the subject of this paper. The structure of this work is as follows: section~\ref{sec:sim} introduces the setup of a GPD detector in simulation, and it also shows some comparisons of the polarization detection performances between Ne and Xe, including detection efficiency and track length. Section~\ref{sec:PolarizationDetection} shows a detailed investigation of the contribution of each bound shell, the angle distribution of emitted photoelectrons using different electromagnetic models, and a modification was found to be necessary for the model named G4LivermorePolarizedPhotoElectricModel. Section~\ref{sec:con} is the conclusion.

\begin{figure}[h]
\centering
    \includegraphics[width=0.5\textwidth]{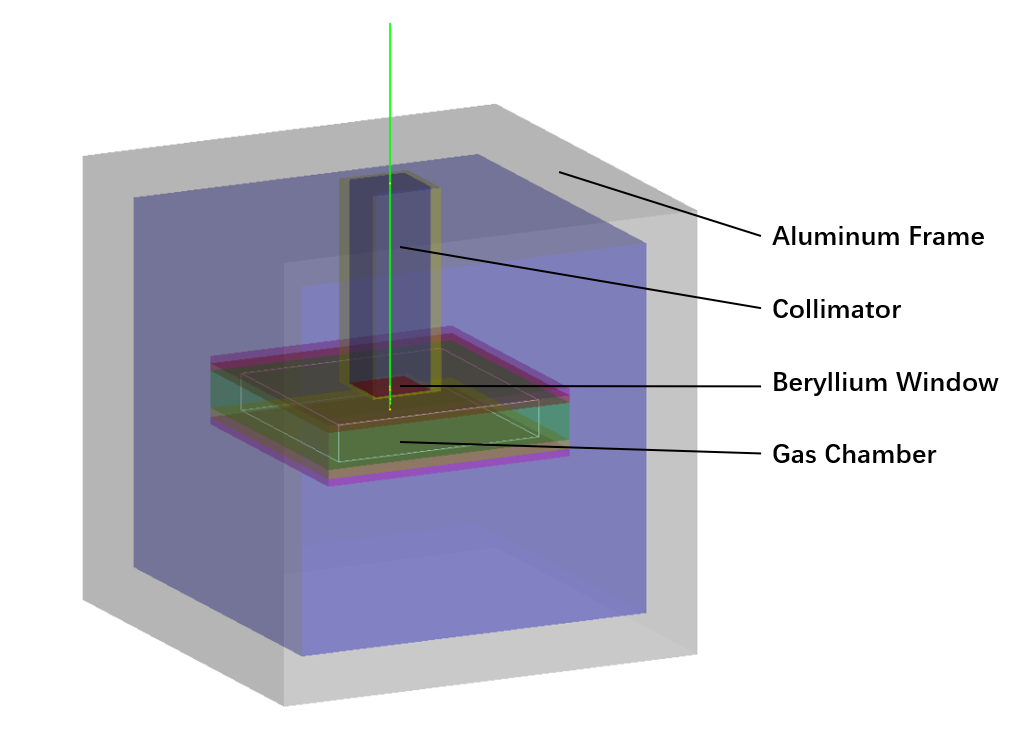}
    \caption{\label{fig:GPD} A simple GPD model in Geant4.}
\end{figure}

\section{Detector simulation and performance}
\label{sec:sim}

To study the features of Xe-based Gas Pixel Detector in simulation, Geant4 framework~\cite{Geant4} (version 10.3) was adopted to build a simple GPD model. The structure of the GPD model is similar to our previous work~\cite{HuangXF2021}. As shown in figure~\ref{fig:GPD}, the main structure of the GPD is a gas chamber made of aluminum, the thickness of gas is 10~mm. On the top of the chamber, a 50~$\upmu$m beryllium window and a collimator made of stainless steel were loaded, respectively. Different working gases, such as Ne, DME, and Xe, can be loaded into the gas chamber for simulation study and comparison. This paper focuses on the GPD using Xe as the main working gas (named Xe-based GPD), and the other one using Ne (named Ne-based GPD) is introduced for comparison. 

\begin{figure}[ht]
\centering % \begin{center}/\end{center} takes some additional vertical space
\includegraphics[width=.4\textwidth]{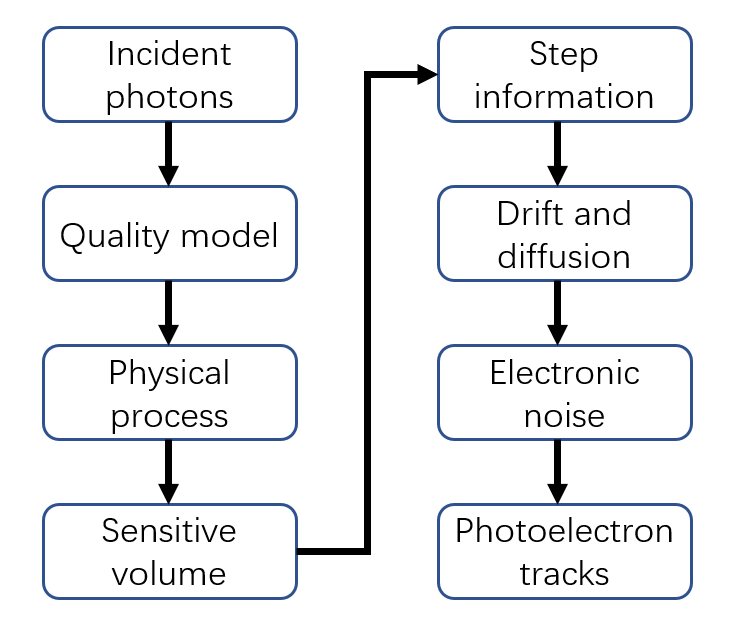}
\caption{\label{fig:SimulationProcess} The process of simulating photoelectron tracks.}
\end{figure}

\begin{figure}[htbp]
\centering
\begin{subfigure}{0.4\textwidth}
\includegraphics[width=0.9\linewidth]{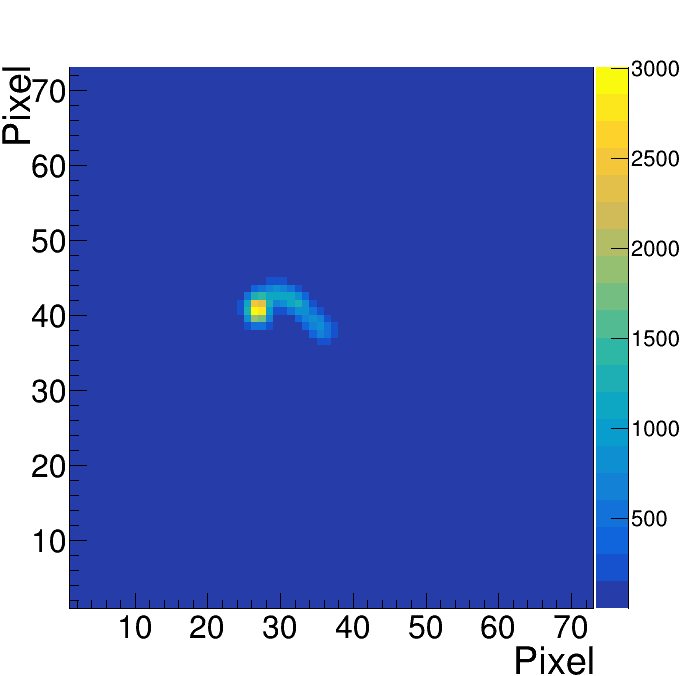}
\caption{\label{fig:Tracks:a} An example of the simulated track in Ne-based GPD.}
\end{subfigure}
\quad
\begin{subfigure}{0.4\textwidth}
\includegraphics[width=0.9\linewidth]{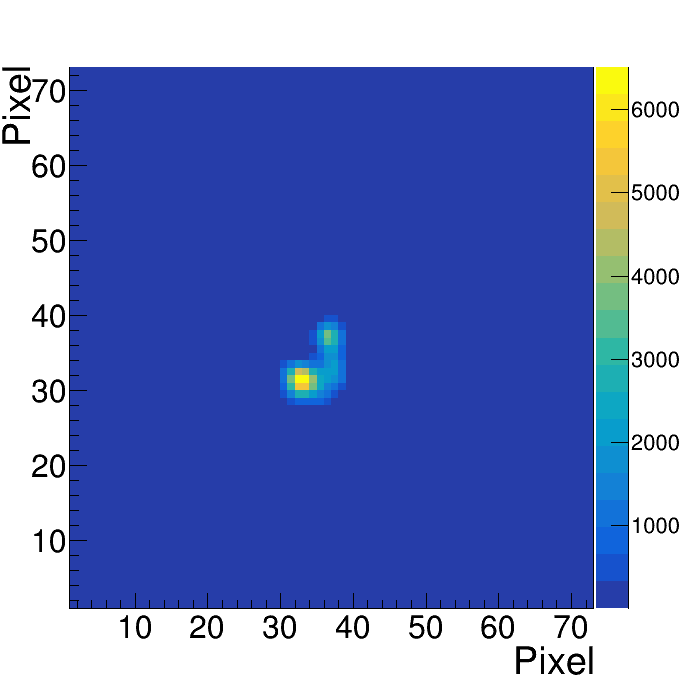}
\caption{\label{fig:Tracks:b} An example of the simulated track in Xe-based GPD.}
\end{subfigure}
\caption{\label{fig:Tracks_good} An example of the simulated track in the Ne-based GPD and the Xe-based GPD. (a) X-rays: 8~keV; working gas: Ne + 20\% DME, 1 atm. (b) X-rays: 20~keV; working gas: Xe + 10\% $\rm CO_{2}$, 1 atm.}
\end{figure}

Figure~\ref{fig:SimulationProcess} shows the flow chart of the photoelectron track simulation. Firstly, X-ray photons of specified energy are injected into the constructed GPD quality model in Geant4, and the photons interact with the passing materials. The detector mainly collects the signals generated by the photoelectric effect between photons and the detection gas. The gas chamber is defined as the sensitive volume to collect simulated information, such as position, energy deposition, and momentum of photoelectrons ejected by the photoelectric effect in each step. The probability of photoelectron emission is calculated from the cross-section of the photoelectric effect, and the direction of emission is calculated from the differential cross-section. The number of secondary electrons of the photoelectron at each step is calculated by dividing the deposition energy by the average ionization energy. In Geant4 simulation, the original tracks of photoelectrons will be recorded immediately. Next, photoelectrons drift in the electric field and ionize the working gas, and then the ionization signal will be amplified several thousands of times through a Gas Electron Multiplier (GEM)~\cite{Bellazzini2013} or a THGEM module~\cite{Breskin2009}. Finally, direct charge sensors such as Topmetal-II$^{-}$~\cite{LiZL2021} can be used for charge collection, resulting in a two-dimensional charge pattern. For simplicity, the process of photoelectrons drift, multiplication and diffusion can be simulated using an effective method, which applying a gaussian diffusion according to the drift distance from the generation location of photoelectron to the collection chip. The $\sigma$ of Gaussian diffusion corresponds to the diffusion coefficient of gas, which can be found in~\cite{Sharma1998} for a variety of gases. Currently, the diffused simulation tracks are projected into a read-out array of 72$\times$72 pixels by considering the setup of our previous experiment~\cite{HuangXF2021}, 0.083~mm$\times$0.083~mm for each pixel. Moreover, a random charge fluctuation will be applied to each pixel to estimate the contribution of electronic noise. To handle transversal diffusion, mixed gas was used according to the experience of theoretical calculation and experiments~\cite{Sharma1998}. In following simulation, the Ne-based GPD and the Xe-based GPD using Ne + 20\% DME and Xe + 10\% CO$_{2}$ as working gas, respectively. And their diffusion coefficients were set to 0.03 $\rm \sqrt{mm}$, and 0.036 $\rm \sqrt{mm}$. With the above operations, the images of the simulated photoelectron tracks can be obtained. Figure~\ref{fig:Tracks_good} shows an example of simulated track in the Ne-based GPD and the Xe-based GPD.

\begin{figure}
\centering
\begin{subfigure}{0.49\textwidth}
\includegraphics[width=0.9\linewidth]{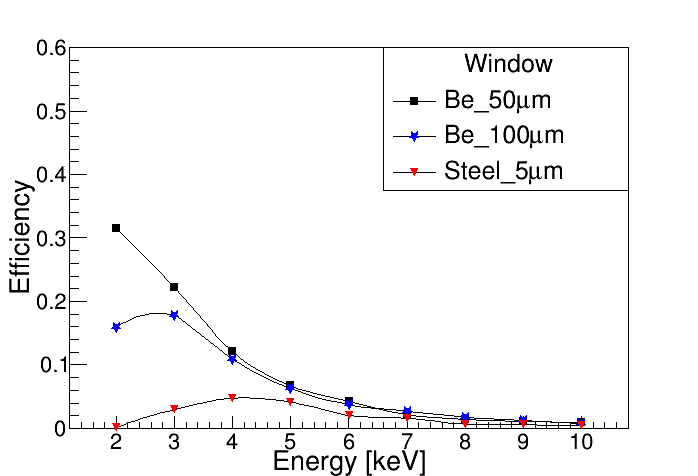}
\caption{\label{fig:Efficiency:a}Ne + 20\% DME.}
\end{subfigure}
\begin{subfigure}{0.49\textwidth}
\includegraphics[width=0.9\linewidth]{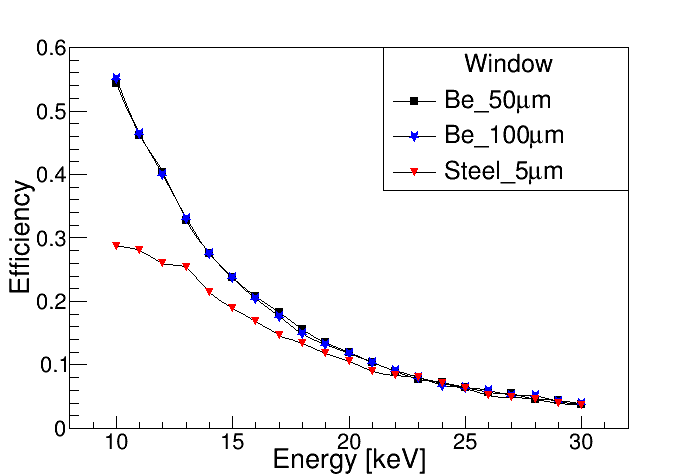}
\caption{\label{fig:Efficiency:b}Xe + 10\% $\rm CO_{2}$.}
\end{subfigure}
\caption{\label{fig:Efficiency}The detection efficiency of the Ne-based, and Xe-based GPD. As the energy increases, the detection efficiency decreases, but the Xe-based GPD still has a high detection efficiency at 10-30~keV. }
\end{figure}

In polarization detection, high detection efficiency is required for the detection of weak X-ray sources, while the long track length and small gas diffusion make reconstruction more efficient and robust. For these reasons, both detection efficiency and track length are investigated in the following. As for the gas diffusion, it mainly depends on the composition of working gas, the strength of the electric field, the gas pressure, and other specific experimental conditions, which need to be further tested and confirmed in future experiments. By calculating the rate of the photoelectric effect in the sensitive volume of the detector, detection efficiency can be defined. Figure~\ref{fig:Efficiency} shows the detection efficiencies of the Ne-based GPD and the Xe-based GPD.  In general, the detection efficiency decreases with the increase of the photon energy for the reason of the cross-section of the photoelectric effect. For the Ne-based GPD, it is difficult to detect the X-ray whose energy is larger than 10~keV. For comparison, the Xe-based GPD has relatively high detection efficiency around 10~keV to 30~keV. The materials and thickness of the window would also affect detection efficiency, especially for the low-energy X-rays whose energy is smaller than 4~keV. With the increase of X-ray energy, the detection efficiency becomes less affected by the thickness of the beryllium window. But there is still a bigger loss of efficiency if denser window material (stainless steel) was used. In the following discussion, a 50~$\upmu$m beryllium window is adopted in default.  

\begin{figure}
\centering
\begin{subfigure}{0.49\textwidth}
\includegraphics[width=0.9\linewidth]{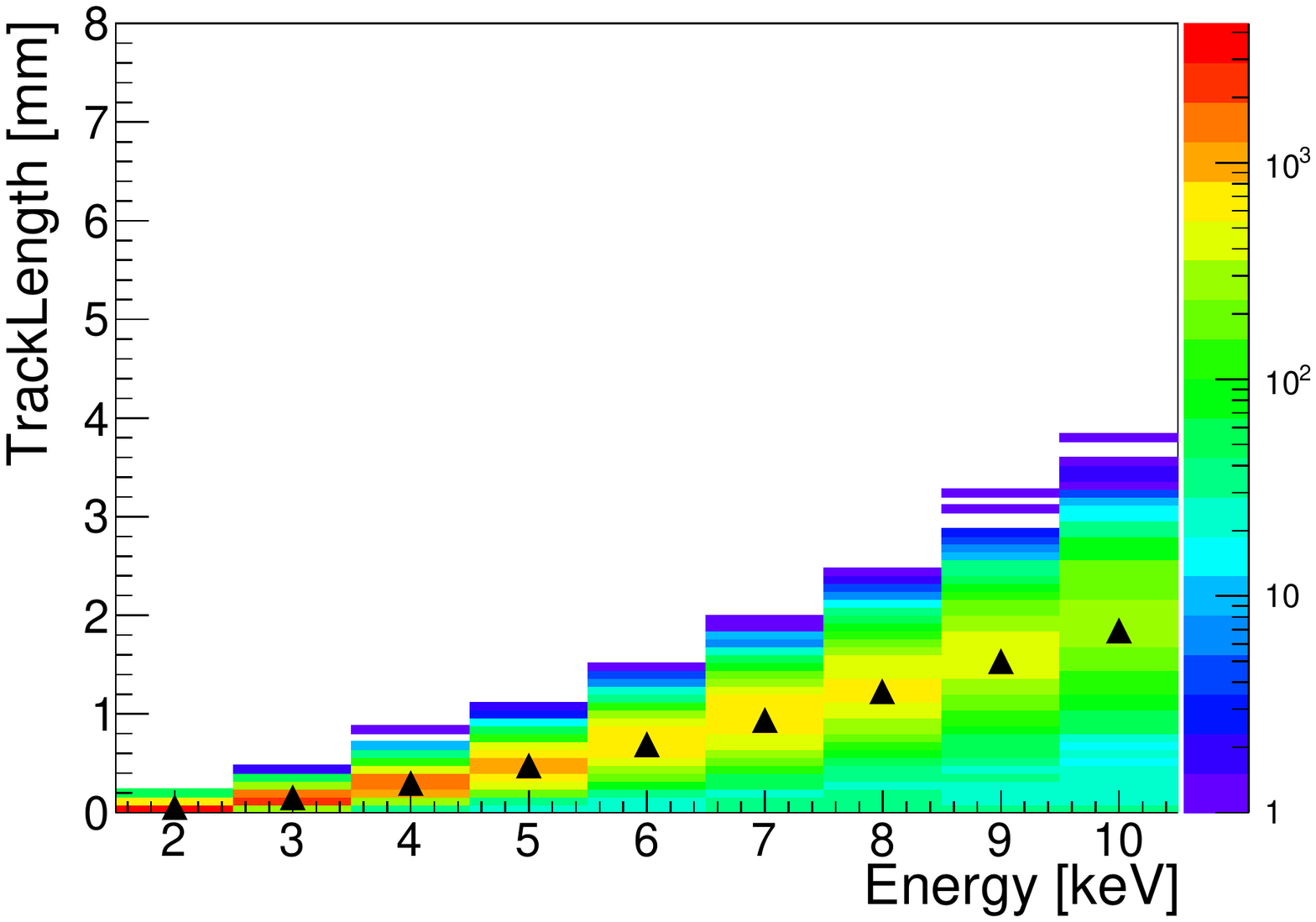}
\caption{\label{fig:ProLen:a}Ne + 20\% DME.}
\end{subfigure}
\begin{subfigure}{0.49\textwidth}
\includegraphics[width=0.9\linewidth]{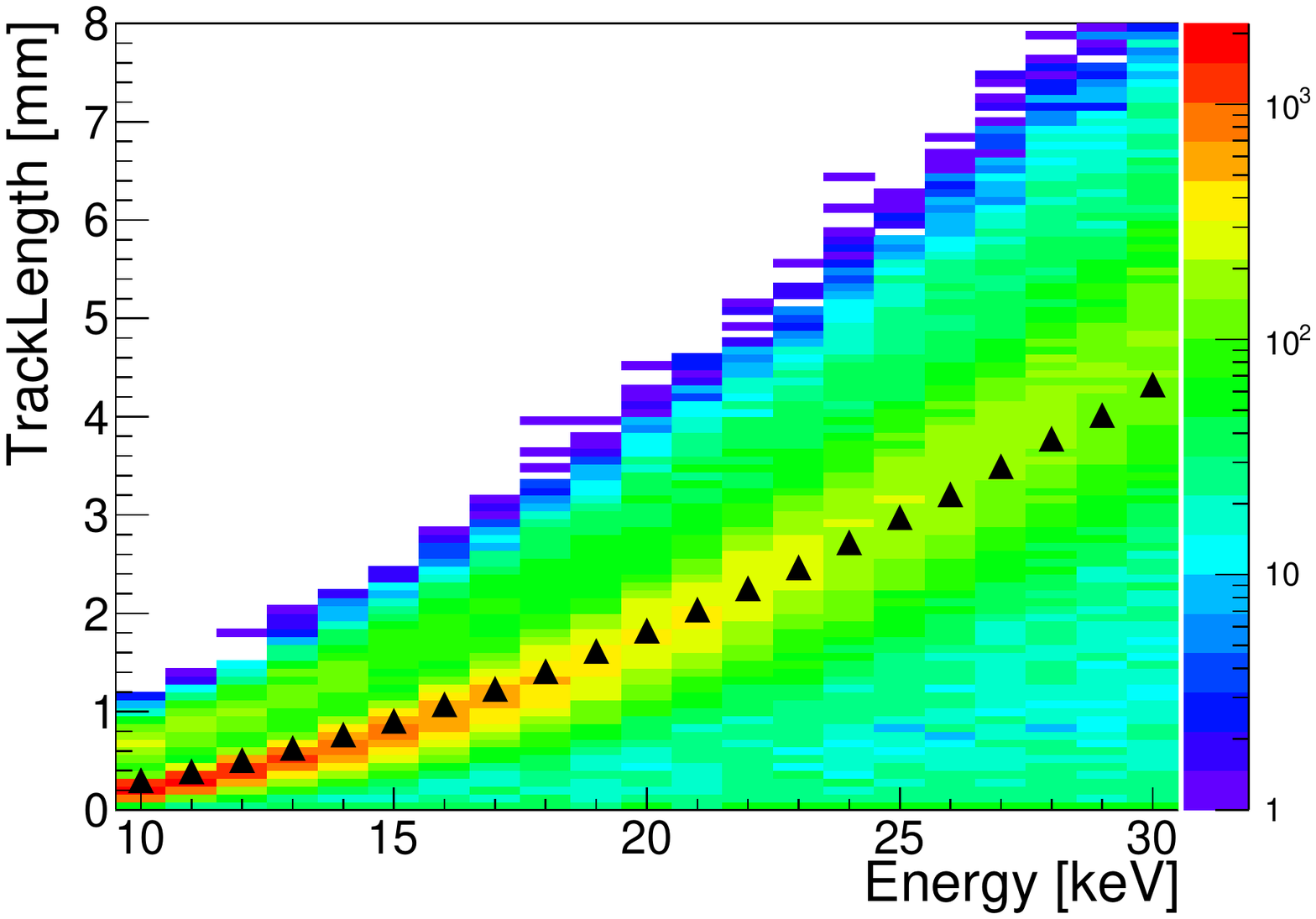}
\caption{\label{fig:ProLen:b}Xe + 10\% $\rm CO_{2}$.}
\end{subfigure}
\caption{\label{fig:ProLen} The distribution of photoelectron track length (on the two-dimensional image) versus X-ray photon energy. (a) The Ne-based GPD; (b) The Xe-based GPD. The black triangle is the average track length in specified X-ray energy.}
\end{figure}

In track reconstruction~\cite{TsingHua3,TsingHua4,LiT2017,HuangXF2021}, high reconstruction efficiency can be obtained for the reconstruction of the long track, but it becomes worse in the case of the short track. As a result, the performance of polarization detection is strongly related to the track length on the output image. Figure~\ref{fig:ProLen} shows the track length distribution of the Ne-based GPD and the Xe-based GPD. Since the average track lengths of photoelectrons are too short for low-energy X-rays (< 4~keV for the Ne-based GPD and < 10~keV for the Xe-based GPD), their reconstruction is difficult and needs further study.   

According to the above analysis, the Xe-based GPD exhibits a good potential in the polarization detection of X-rays at 10-30~keV. Based on the fact that different working gases are suitable for the detection of different energy regions, in a real detector, multiple GPD modules using Ne-based gas and Xe-based gas, respectively, can be integrated into a detection array to achieve a full detection coverage in a 2-30~keV energy range. In addition, the detection array is also useful to obtain a large detection area and it can improve the sensitivity of weak X-ray source detection. 

\section{Polarization detection}
\label{sec:PolarizationDetection}

\begin{figure}[h]
\centering % \begin{center}/\end{center} takes some additional vertical space
\includegraphics[width=.4\textwidth]{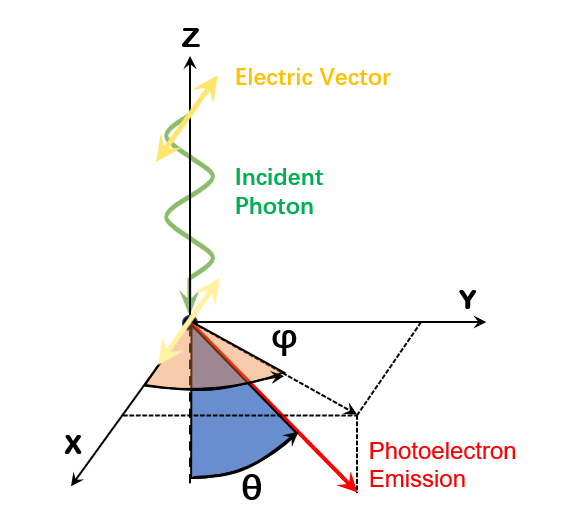}
\caption{\label{fig:Photoelectron} Schematic illustration of polarization detection for X-rays through the photoelectric effect. The exit direction of the photoelectron is mainly described by two angles, whereas the angle between the photoelectron and the incident photon is assigned as $\theta$, and the angle between the projection vector in the XY plane and the X-axis is described as $\varphi$.}
\end{figure}

The principle of polarization detection for X-rays through the photoelectric effect is shown in figure~\ref{fig:Photoelectron}. When an incident X-ray photon interacts with working gas through the photoelectric effect, the photon will be absorbed and a photoelectron is ejected by the absorber atom from one of its bound shells~\cite{RadiationBook}. The kinetic energy of the emitted photoelectron depends on the energy of the incident photon and the binding energy of the photoelectron in its original shell. And the $\varphi$ angle (figure~\ref{fig:Photoelectron}) of the emitted photoelectron carries the polarization information of the incident X-rays. In this section, to further understand the characteristics and features of the Ne-based and Xe-based GPDs, a detailed investigation of the contribution of each bound shell and the emission angle distribution of photoelectrons will be presented. To highlight the main components of the working gas (Ne and Xe) that dominate the detection, pure Ne and pure Xe will be used in the following simulations. This arrangement will not bias the conclusion. 

\subsection{The contribution of each bound shell}
\label{subsec:Shell}

The kinetic energy ($E_{p.e.}$) of the emitted photoelectron is given by the following expression$\colon$

\begin{equation}
\label{eq:BindingEnergy}
E_{p.e.} = E_{\gamma} - E_{binding}\left( Z_{i} \right)
\end{equation}

Where $E_{\gamma}$ is the energy of the incident photon, and $E_{binding}\left( Z_{i} \right)$ represents the electron binding energy of different shells, which is listed in the database of G4AtomicShells (Table~\ref{tab:BindingEnergy})~\cite{Carlson1976}. Considering the differences in electron binding energies, the emitted photoelectrons can be classified into different types according to the shell layers they come from. 

\begin{table}[!ht]
    \caption{\label{tab:BindingEnergy} Electron binding energies (eV) of Ne and Xe.}
    \centering
    \begin{threeparttable}        
      \begin{tabular}{*9{c}}\toprule
        Element & K $1s$ & $L_{1}$ $2s$ & $L_{2}$ $2p_{1/2}$ & $L_{3}$ $2p_{3/2}$ & $M_{1}$ $3s$ & $M_{2}$ $3p_{1/2}$ & $M_{3}$ $3p_{3/2}$ & $M_{4}$ $3d_{3/2}$ \\ \midrule
        ${}_{10}^{20}Ne$ & 870.1 & 48.47 & 21.66 & 21.56\\[3pt]
        ${}_{54}^{131}Xe$\tnote{1} & 34570. & 5460. & 5110. & 4790. & 1148.7 & 1002.1 & 940.6 & 689.0\\[3pt] \bottomrule
      \end{tabular}
         \begin{tablenotes}   
        \footnotesize              
        \item[1] The higher shells in Xe are not listed. 
      \end{tablenotes}          
    \end{threeparttable}      
\end{table}

\begin{figure}
\centering
\begin{subfigure}{0.45\textwidth}
\includegraphics[width=0.9\textwidth]{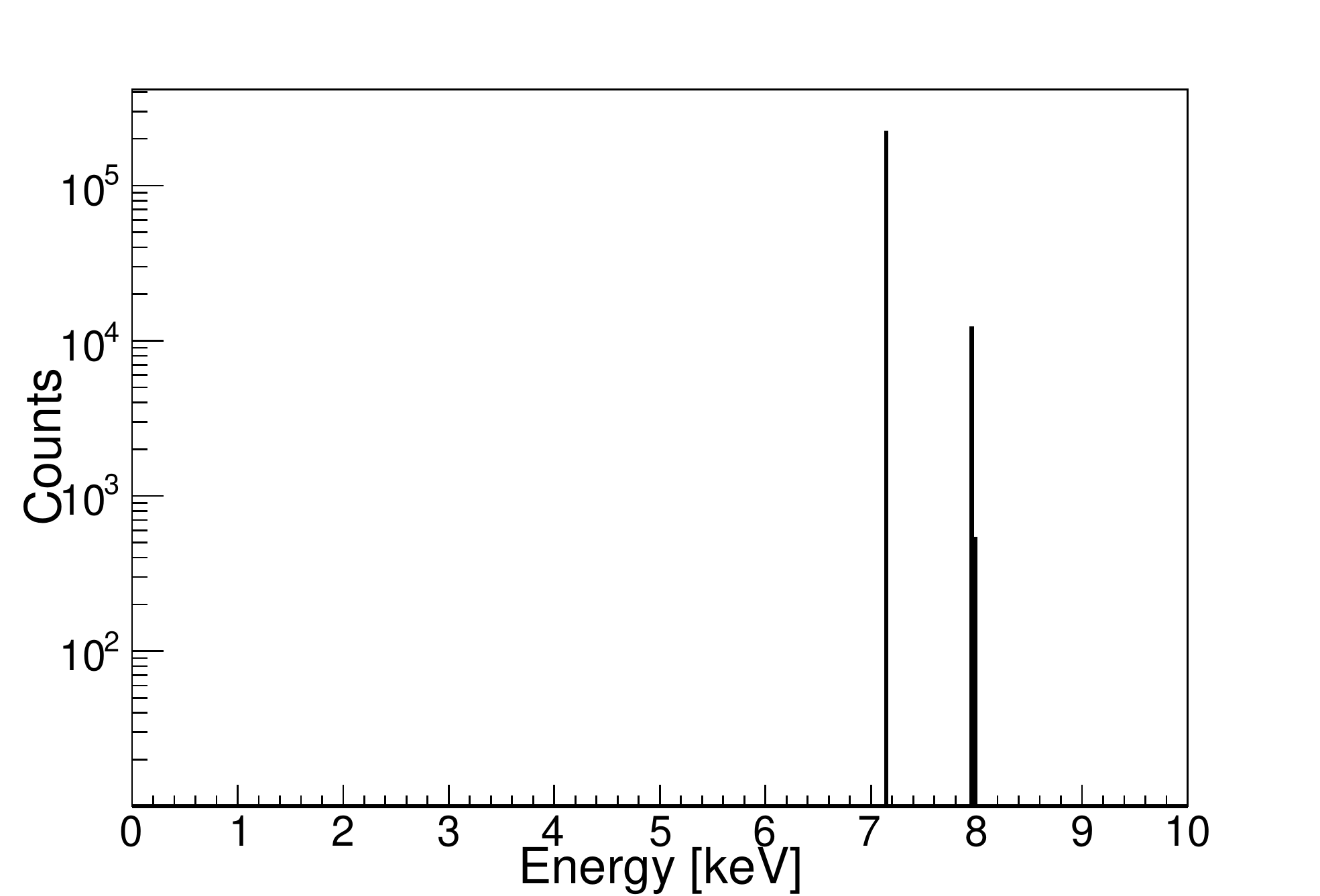}
\caption{\label{fig:KinEnergy:Ne} The kinetic energy distribution of the emitted photoelectrons, in the case of injecting 8~keV photons into the GPD using pure Ne as working gas. }
\end{subfigure}
\quad
\begin{subfigure}{0.45\textwidth}
\includegraphics[width=0.9\textwidth]{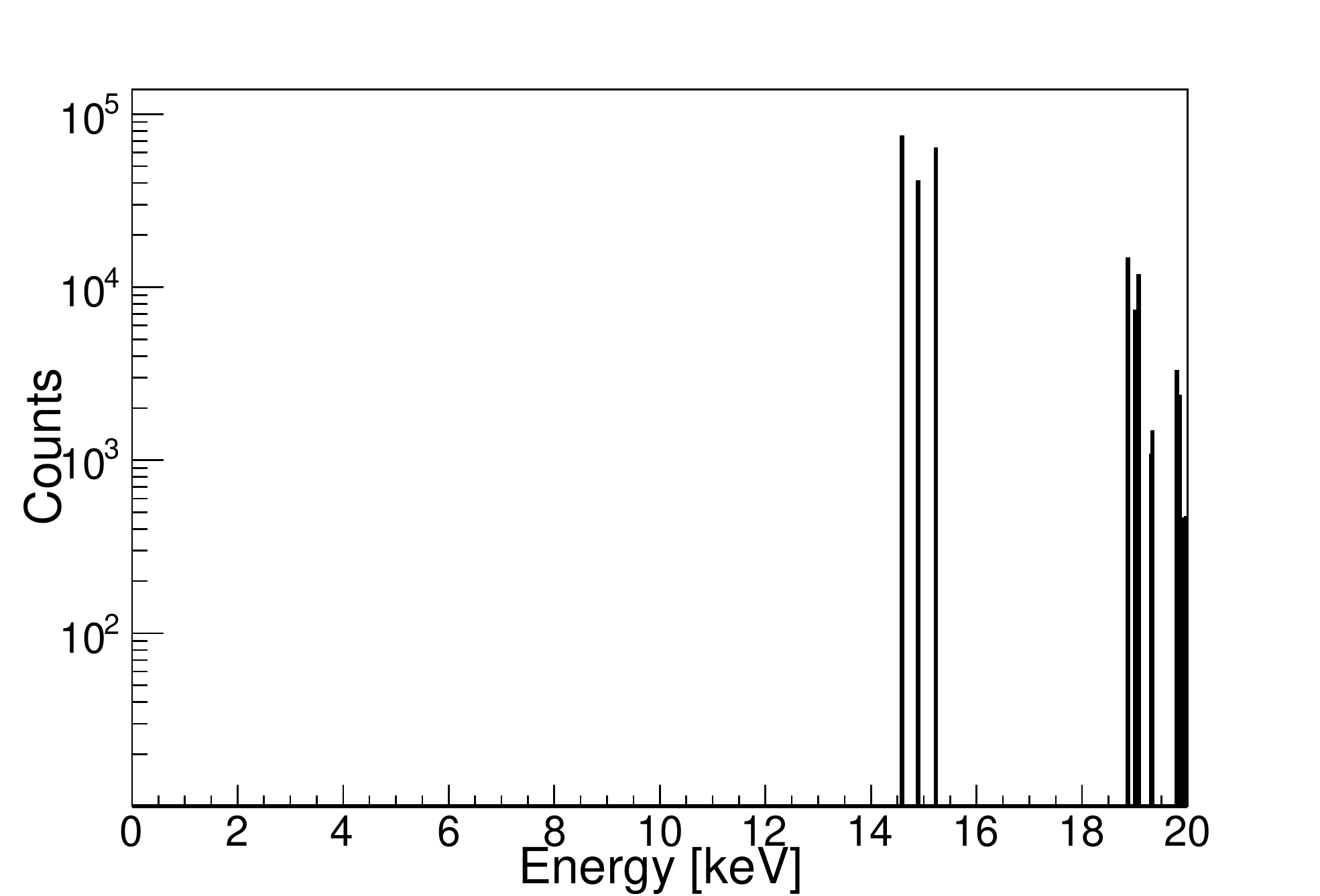}
\caption{\label{fig:KinEnergy:Xe} The kinetic energy distribution of the emitted photoelectrons, in the case of injecting 20~keV photons into the GPD using pure Xe as working gas. }
\end{subfigure}
\caption{\label{fig:KinEnergy}The kinetic energy distribution of the emitted photoelectrons. These peaks correspond to the photoelectrons that come from different bound shells. (a) The highest peak is from the K shell, three peaks (L1, L2, L3) around 8~keV overlap together caused by the small differences between their binding energies; (b) From left to right, the peaks correspond to the L1 shell, L2 shell, L3 shell, and several M shells, respectively.}
\end{figure}

\begin{figure}
\centering
\begin{subfigure}{0.45\textwidth}
\includegraphics[width=0.9\textwidth]{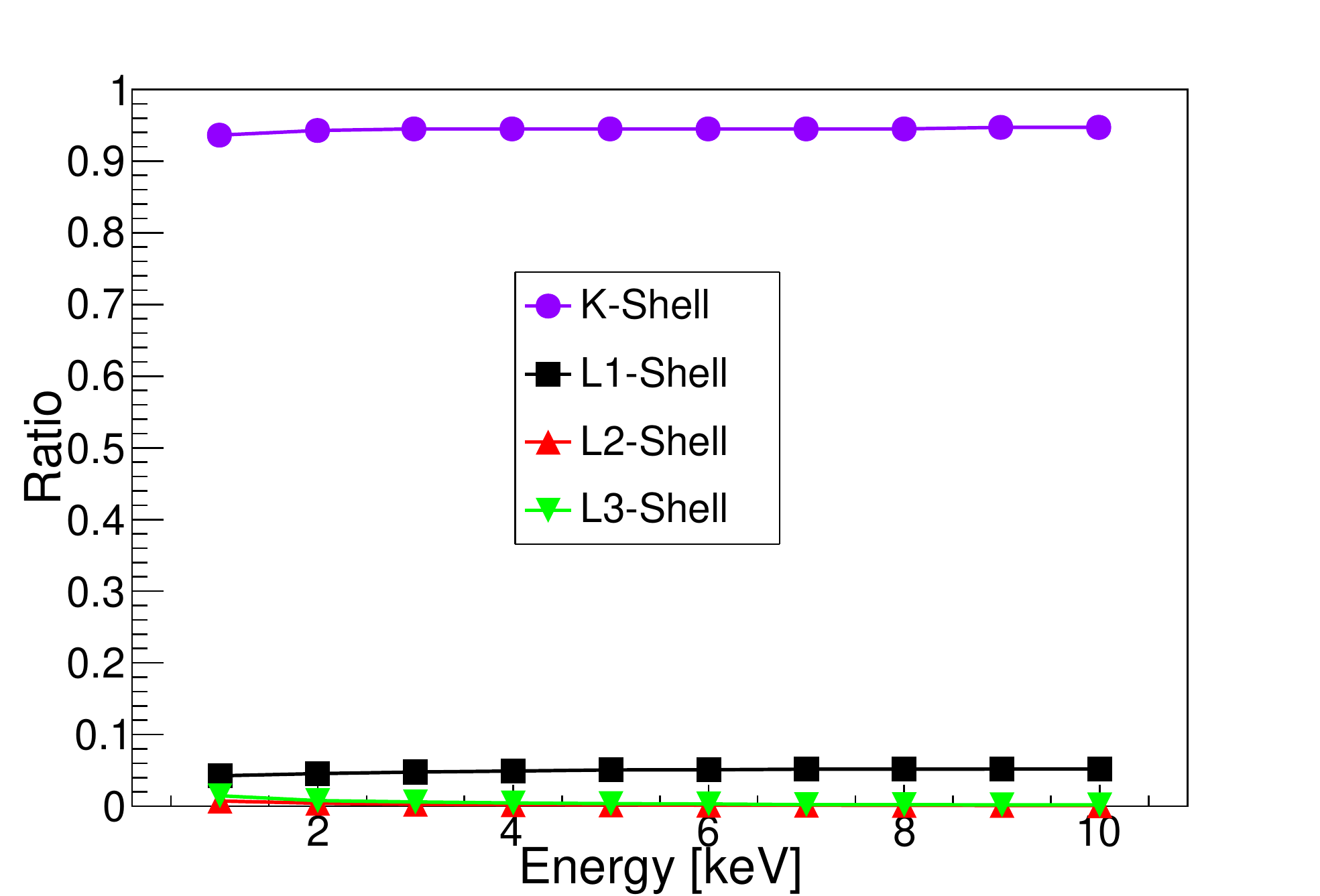}
\caption{The photoelectrons contribution from each shell in the case of using pure Ne for detection. }
\end{subfigure}
\quad
\begin{subfigure}{0.45\textwidth}
\includegraphics[width=0.9\textwidth]{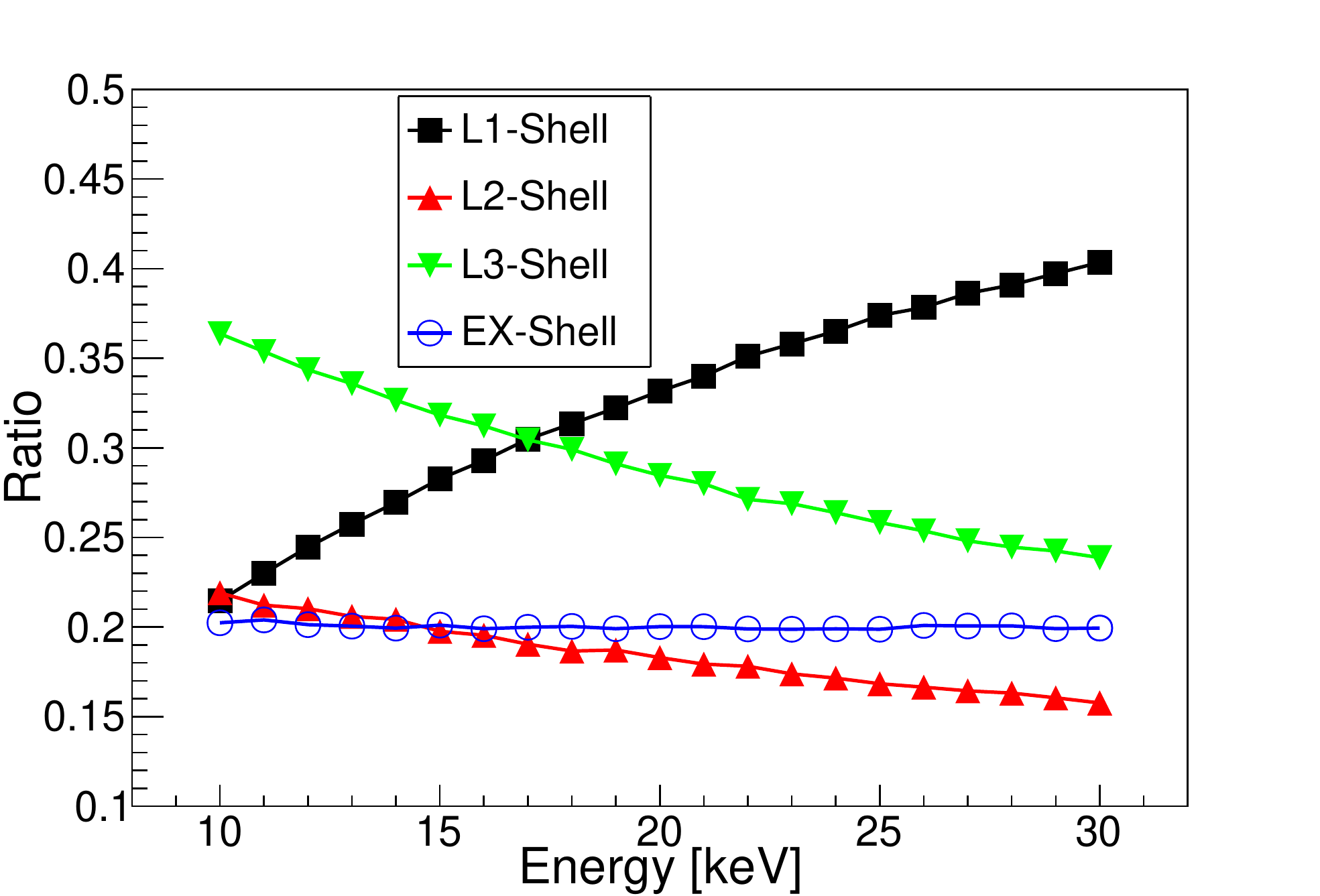}
\caption{The photoelectrons contribution from each shell in the case of using pure Xe for detection.}
\end{subfigure}
\caption{\label{fig:ShellRatio}The contribution from photoelectrons emitted by each shell. (a) The K shell photoelectrons (purple circles) dominate in the detection at 2-10~keV. And the contribution of other shells is less than 6\% (L1: black squares; L2: red triangles; L3: green inverted triangles). (b) Photoelectrons from three L shells dominate in the detection at 10-30~keV (L1 to L3: the same labels as in figure (a)). The remaining higher shells are integrated and labeled as the EX shell (blue hollow circles).}
\end{figure}

The kinetic energy distribution of the emitted photoelectrons can be found in figure~\ref{fig:KinEnergy}. Figure~\ref{fig:KinEnergy:Ne} corresponds to the detection of 8~keV photons in pure Ne, four peaks can be found and they come from the K shell, L1 shell, L2 shell, and L3 shell, respectively. Figure~\ref{fig:KinEnergy:Xe} is the detection of 20~keV photons in pure Xe, from left to right, the peaks correspond to the L1 shell, L2 shell, L3 shell, and several M shells, respectively. As shown in figure~\ref{fig:ShellRatio}, for Ne, the K shell photoelectrons dominate in the detection at 2-10~keV. As for Xe, since the electron binding energy of the K shell is higher than 30~keV, the K shell photoelectrons cannot be emitted when the energy of the detected X-rays is less than 30~keV. With the increase of energy, the ratio of emitted photoelectrons from the L2 and L3 shells progressively decreases, while the ratio from the L1 shell gradually increases, yielding the ratio of the whole L shell (L1 + L2 + L3) of about 80\%. Therefore, the feature of the photoelectrons in the L shell is considered as the key point for investigating X-ray polarization detection using Xe.

\subsection{The distribution of emission angle}
\label{subsec:Angle}

In polarization detection, the polarization information of the detected X-ray source can be calculated using the emission angle distribution of photoelectrons. Next, details about photoelectron emission will be introduced. 

In general, when a photoelectron emits through the photoelectric effect, its emission angle is sampled according to the differential cross-section of reaction using the calculation kernel in Geant4. There are multiple electromagnetic models available in Geant4 for the study of photoelectric physics. Table~\ref{tab:DiffModel} summarizes the details of six commonly used electromagnetic models in Geant4. In these models, the G4LivermorePolarizedPhotoElectricModel (named LM model) and the G4LivermorePolarizedPhotoElectricGDModel (named GD model) were found to include both the shell structure and polarization information of emitted photoelectrons~\cite{Geant4manual}. Hence, further investigations were carried out based on the LM model and GD model. 

\begin{table}[!htbp]
    \caption{\label{tab:DiffModel} Comparison of different electromagnetic models in Geant4.}
    \centering
    \begin{threeparttable}        
      \begin{tabular}{*3{c}}\toprule
       \multirow{2}{*}{Model} & \multicolumn{2}{c}{Behaviors of emitted photoelectron}\\ 
\cline{2-3}    & \makecell*[c]{With shells \\structure}       & \makecell*[c]{With polarization \\information} \\ \midrule
        G4PhotoElectricEffect                           & -                       & -\\
        G4PolarizedPhotoElectricEffect                  & -                       & -\\ 
        G4LivermorePhotoElectricModel                   & $\surd$                 & -\\ 
        G4LivermorePolarizedPhotoElectricModel          & $\surd$                 & $\surd$\\ 
        G4PenelopePhotoElectricModel                    & $\surd$                 & -\\ 
        G4LivermorePolarizedPhotoElectricGDModel        & $\surd$                 & $\surd$\\  \bottomrule
      \end{tabular}
         \begin{tablenotes}   
        \footnotesize              
        \item[1] "-" means undetected. 
      \end{tablenotes}          
    \end{threeparttable}      
  \end{table}

\subsubsection{The LM model}
\label{sec:LM model}
In the LM model, the differential cross-section of the K shell is provided by the following equation~\cite{Gavrila1959}$\colon$

\begin{equation}\label{eq:GavilaKshell}
\begin{gathered}
\frac{d \sigma_{K}}{d \omega}(\theta, \varphi)=\frac{4}{m^{2}} \alpha^{6} Z^{5} \frac{\beta^{3}\left(1-\beta^{2}\right)^{3}}{\left[1-\left(1-\beta^{2}\right)^{1 / 2}\right]}\left(F\left(1-\frac{\pi \alpha Z}{\beta}\right)+\pi \alpha Z G\right) \\
F=\frac{\sin ^{2} \theta \cos ^{2} \varphi}{(1-\beta \cos \theta)^{4}}-\frac{1-\left(1-\beta^{2}\right)^{1 / 2}}{2\left(1-\beta^{2}\right)} \frac{\sin ^{2} \theta \cos ^{2} \varphi}{(1-\beta \cos \theta)^{3}} \\
+\frac{\left[1-\left(1-\beta^{2}\right)^{1 / 2}\right]^{2}}{4\left(1-\beta^{2}\right)^{3 / 2}} \frac{\sin ^{2} \theta}{(1-\beta \cos \theta)^{3}} \\
G=\frac{\left[1-\left(1-\beta^{2}\right)^{1 / 2}\right]^{1 / 2}}{2^{7 / 2} \beta^{2}(1-\beta \cos \theta)^{5 / 2}}\left[\frac{4 \beta^{2}}{\left(1-\beta^{2}\right)^{1 / 2}} \frac{\sin ^{2} \theta \cos ^{2} \varphi}{1-\beta \cos \theta}+\frac{4 \beta}{1-\beta^{2}} \cos \theta \cos ^{2} \varphi \right. \\
-4 \frac{1-\left(1-\beta^{2}\right)^{1 / 2}}{1-\beta^{2}}\left(1-\cos ^{2} \varphi\right)-\beta^{2} \frac{1-\left(1-\beta^{2}\right)^{1 / 2}}{1-\beta^{2}} \frac{\sin ^{2} \theta}{1-\beta \cos \theta} \\
\left.+4 \beta^{2} \frac{1-\left(1-\beta^{2}\right)^{1 / 2}}{\left(1-\beta^{2}\right)^{3 / 2}}-4 \beta \frac{\left[1-\left(1-\beta^{2}\right)^{1 / 2}\right]^{2}}{\left(1-\beta^{2}\right)^{3 / 2}}\right] \\
+\frac{1-\left(1-\beta^{2}\right)^{1 / 2}}{4 \beta^{2}(1-\beta \cos \theta)^{2}}\left[\frac{\beta}{1-\beta^{2}}-\frac{2}{1-\beta^{2}} \cos \theta \cos ^{2} \varphi+\frac{1-\left(1-\beta^{2}\right)^{1 / 2}}{\left(1-\beta^{2}\right)^{3 / 2}} \cos \theta\right. \\
\left.-\beta \frac{1-\left(1-\beta^{2}\right)^{1 / 2}}{\left(1-\beta^{2}\right)^{3 / 2}}\right]
\end{gathered}
\end{equation}

where $\beta$ is the velocity of the photoelectron, $\alpha$ is the fine structure constant, Z is the atomic number of the material, $\varphi$ and $\theta$ are the photoelectron emission angles (figure~\ref{fig:Photoelectron}). In the case of first-order correction of $\alpha Z$, the differential cross-section of L1 shell~\cite{Gavrila1961} is approximate as follows$\colon$

\begin{equation}
\label{eq:GavilaLshell}
d\sigma_{L1} = \zeta\frac{1}{8}d\sigma_{K}
\end{equation}

where $d\sigma_{K}$ and $d\sigma_{L1}$ are the differential cross-section of K and L1 shells, respectively. As mentioned in~\cite{Gavrila1961}, the parameter $\zeta$ is equal to 1 (then, $d\sigma_{L1} = \frac{1}{8}d\sigma_{K}$) when working with unscreened Coulomb wave functions, and it has been proved that this approximation works well for both low Z and medium Z. However, it was found that $d\sigma_{L1} = d\sigma_{K}$ was applied in the LM model according to our investigations, which is different to what it claims. As a result, this operation overestimates the differential cross-sections of L shells when sampling the emission angle and it introduces a nonphysical result of the emission angle distribution which will be discussed later (section~\ref{sec:true}). On the other hand, the L1 shell differential cross-section is also employed for higher shells in the LM model.

\subsubsection{The GD model}

The employed GD model does not calculate the differential cross-section of each shell separately, but all shells share the same calculation of differential cross-section according to the following expression~\cite{Depaola2006} $\colon$

\begin{equation}
\label{eq:Depaola}
d\sigma \propto \frac{\sin^{2}\theta}{\left( {1 - \beta{\mathit{\cos}\theta}} \right)^{4}}\left\{ {\frac{k^{2}}{4}\left\{ {1 - \beta{\mathit{\cos}\theta}} \right\} + \left\lbrack {\frac{1}{\varepsilon} - \frac{k}{2}\left\lbrack {1 - \beta{\mathit{\cos}\theta}} \right\rbrack} \right\rbrack{\cos^{2}\varphi}} \right\}
\end{equation}

where $\varepsilon$ is the total energy of the electron, and $k$ is the total energy of the photon. In the GD model, all shells follow this differential cross-section distribution and then emit photoelectrons according to different shell energies.

\subsubsection{Model comparison}
\label{sec:true}

The differential cross-section calculation of the above two models demonstrates that the $\varphi$ angular distribution of each bound shell is approximate to cosine square, which is:

\begin{equation}
\label{eq:cosine}
d\sigma \propto {\cos^{2}\varphi}
\end{equation}

\begin{figure}[h]
\centering % \begin{center}/\end{center} takes some additional vertical space
\includegraphics[width=.9\textwidth]{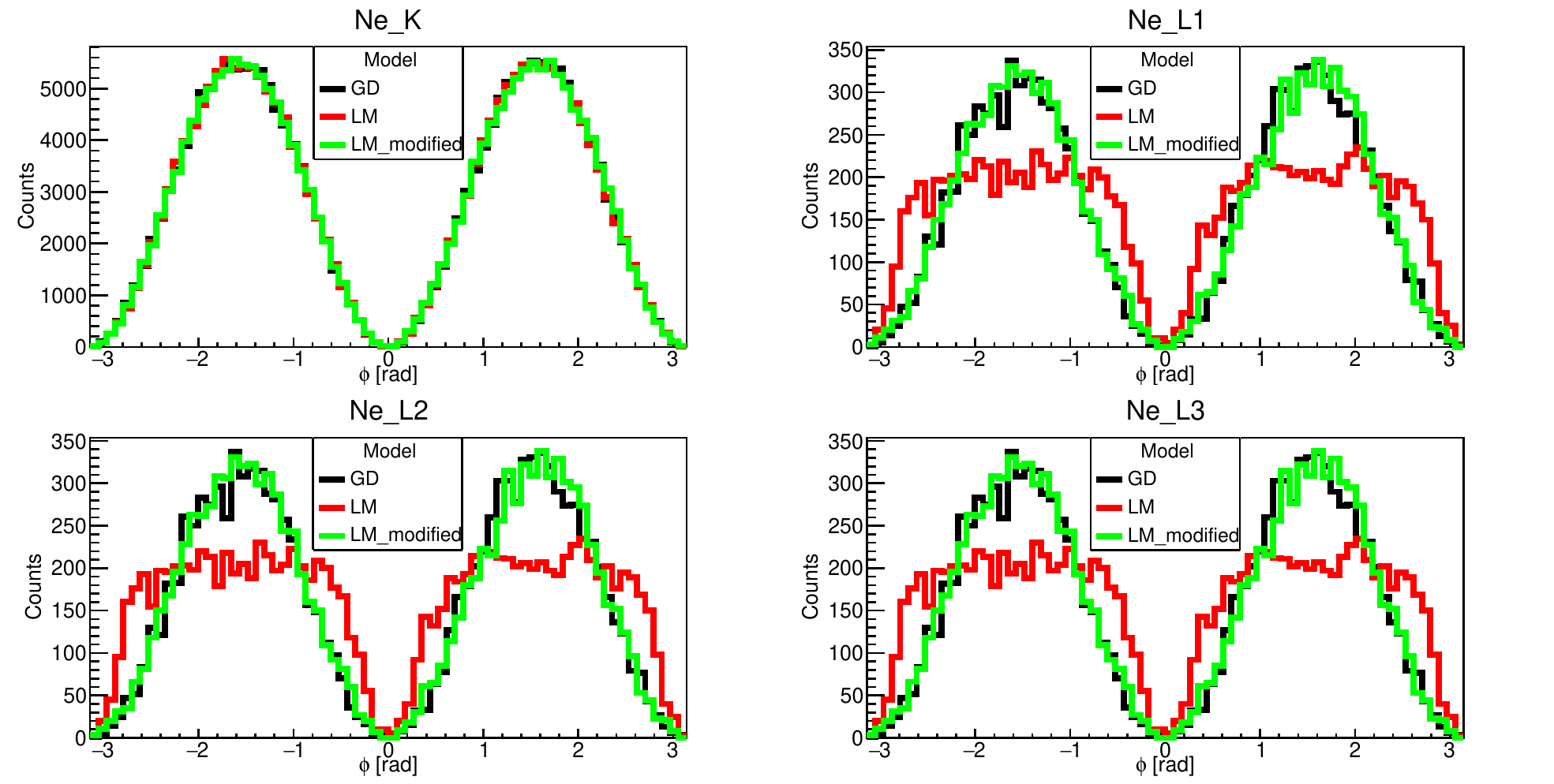}
\caption{\label{fig:NeDisstribution} The emission angle ($\varphi$) distribution of each bound shell (K, L1, L2, L3), in the case of injecting 8~keV linear polarization photons into the GPD using pure Ne as working gas. The simulation results of L shells using the LM model (red color) have serious distortions and their distributions are no longer a $\cos^{2}\varphi$ shape. The simulation results of the GD model and the LM\_modified model are in black and green color, respectively.}
\end{figure}

\begin{figure}[h]
\centering % \begin{center}/\end{center} takes some additional vertical space
\includegraphics[width=.9\textwidth]{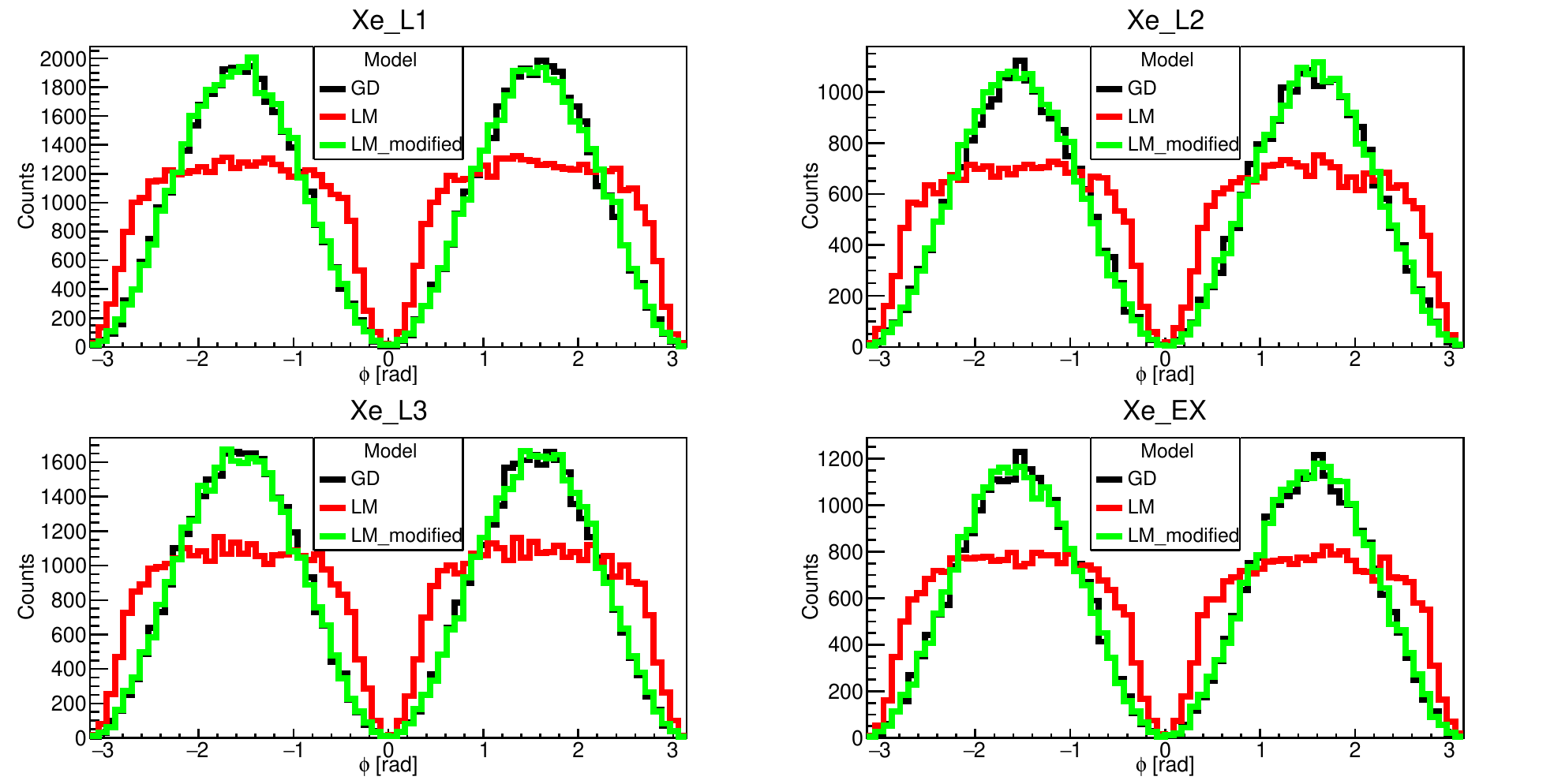}
\caption{\label{fig:XeDisstribution} The emission angle ($\varphi$) distribution of each bound shell (L1, L2, L3 and others shells), in the case of injecting 20~keV linear polarization photons into the GPD using pure Xe as working gas. The simulation results of L shells and higher shells using the LM model (red color) have serious distortions and their distributions are no longer a $\cos^{2}\varphi$ shape. The simulation results of the GD model and the LM\_modified model are in black and green color, respectively.}
\end{figure}

Moreover, the calculation in \cite{Depaola2006} (referenced by the GD model) derived the same calculation result as Sauter~\cite{Sauter1955} in K shell, while the calculation in \cite{Gavrila1961} (referenced by the LM model) extended Sauter's calculation to obtain a higher order of approximation for the calculation of the differential cross-section, and it acquires the same calculation result as Pratt~\cite{Pratt1960} in L1 shell. According to the above discussions, the two models should have consistent theoretical predictions for the emission angle distribution of photoelectrons. However, as shown in figure~\ref{fig:NeDisstribution} and figure~\ref{fig:XeDisstribution}, which are the emission angle distribution of each bound shell in Ne and Xe using linear polarization photons as the incident X-ray source, their distributions of L shells are different between the LM model (red color) and the GD model (black color). The expected shape ($\cos^{2}\varphi$) has serious distortion in the simulation result using the LM model. This effect is because the LM model overestimates the differential cross-sections of L shells (section~\ref{sec:LM model}) when sampling the emission angle. As a result, the top part of the normal distribution will be discarded when sampling the emission angle using the acceptance-rejection technique~\cite{Geant4manual,2BNbremsstrahlug} in the LM model.

To validate the emission angle distribution in simulation, we modify the LM model by including the missing constant $\zeta\frac{1}{8}$ when calculating the differential cross-sections of L shells, the modified model named as LM\_{modified} model in the following. After modification, the simulation results of the LM\_modified model (green color) are in good agreement with the GD model (black color), as shown in figure~\ref{fig:NeDisstribution} and figure~\ref{fig:XeDisstribution}. This consistency not only validates the modification, but also provides an important reference for the simulation study of polarimeter in the future.

\subsection{Modulation factor comparison}
\label{sec:ModulationComparison}

When a photoelectron is detected and its two-dimensional image (figure~\ref{fig:Tracks_good}) is available from read-out electronics, to calculate polarization information, a track reconstruction algorithm is required to get the original emission direction of the photoelectron. In general, two kinds of algorithms have been developed and successfully applied. One of them was developed based on moment analysis~\cite{TsingHua3,TsingHua4,HuangXF2021}, the other one based on the shortest path problem in graph theory~\cite{LiT2017}. In our study, a similar reconstruction algorithm as in~\cite{LiT2017} was applied to reconstruct the simulated tracks, which were generated based on the process mentioned in section~\ref{sec:sim}. 

\begin{figure}[h]
\centering
\includegraphics[width=0.5\textwidth]{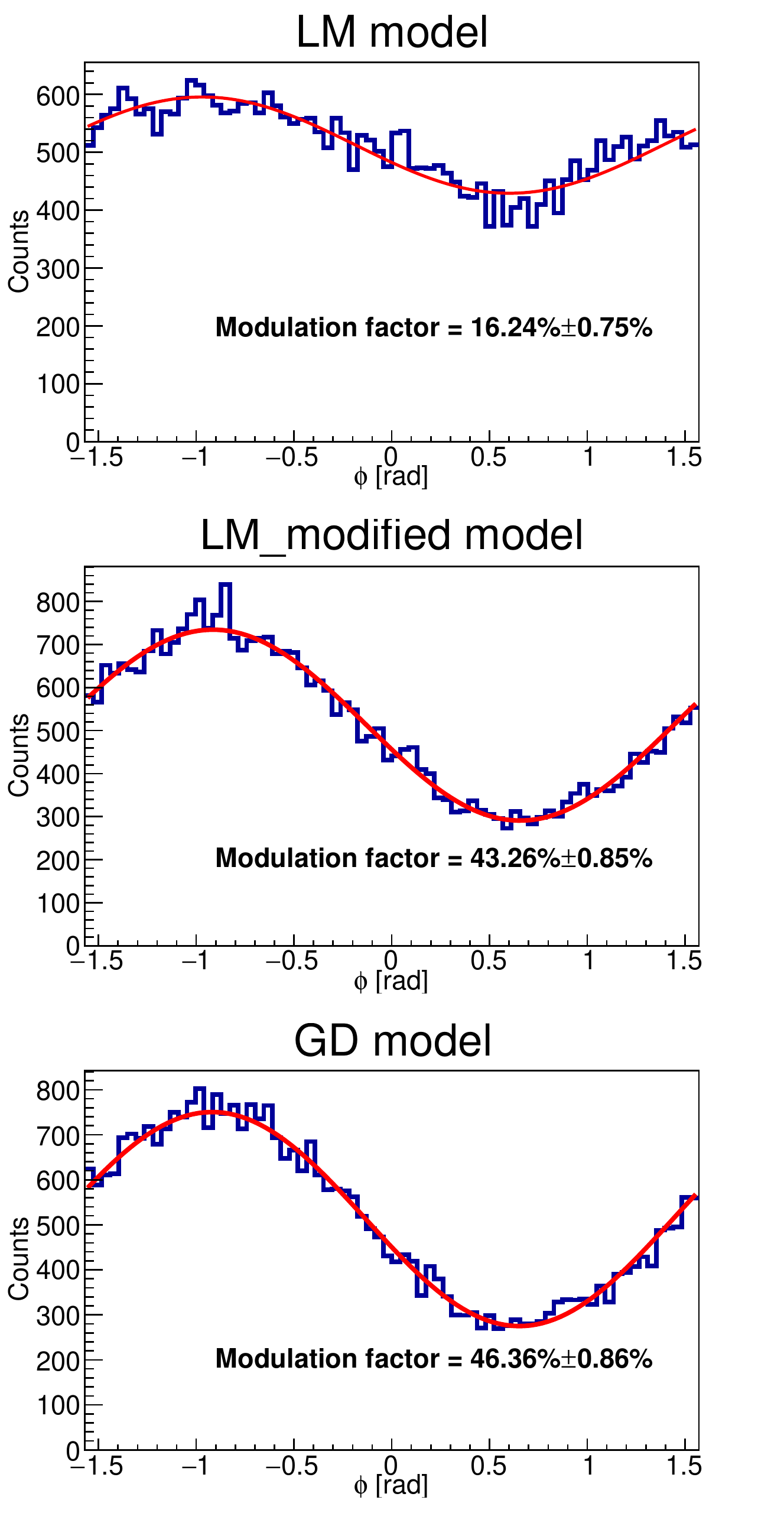}
\caption{\label{fig:Reconstruction} Reconstruction results of 20~keV photoelectron tracks simulated by different models.}
\end{figure}

Figure~\ref{fig:Reconstruction} shows the reconstructed emission angle distribution, 20~keV linear polarization photons were detected by the Xe-based GPD. The simulated tracks were generated using the LM model, the LM\_modified model, and the GD model, respectively. Compared to the ideal distribution, the reconstructed distributions include a constant component caused by the limitation of detection and the precision of reconstruction. Moreover, the reconstructed distributions of the LM model are flatter than the other two models. It is because the LM model overestimates the differential cross-sections of L shells when sampling the emission angle (section~\ref{sec:true}). Modulation factor is a parameter to estimate the ability of polarization detection, which can be calculated using the maximum and
minimum values of the reconstructed distribution. In figure~\ref{fig:Reconstruction}, the reconstructed distributions were fitted by~\eqref{eq:fitfunction} to obtain their modulation factors~\eqref{eq:modulation}. A and B correspond to the nonpolarized component and the polarized component, respectively.

The modulation factor is 16.24\% for the simulation using the LM model, while it becomes much better in case using the LM\_modified model and the GD model, which is 43.26\% and 46.36\%, respectively. Once again, the improvement of the modulation factor indicates that our modification is necessary to the simulation study of polarization detection. 

\begin{equation}
\label{eq:fitfunction}
M(\varphi) = A + B\cos^{2}(\varphi-\varphi_{0})
\end{equation}

\begin{equation}
\label{eq:modulation}
\mu=\frac{M_{\mathrm{Max}}-M_{\mathrm{Min}}}{M_{\mathrm{Max}}+M_{\mathrm{Min}}}=\frac{B}{2 A+B}
\end{equation}

\section{Conclusion}
\label{sec:con}

In this work, a Xe-based gas pixel detector was constructed in simulation, and it shows good performances and potential in the polarization detection of X-rays at 10-30~keV. And the photoelectrons are mainly contributed by the L shells since the binding energy of the K shell is larger than 30~keV. Furthermore, different electromagnetic models are investigated to verify the distribution of emission angle, and it was found that the G4LivermorePolarizedPhotoElectricModel in Geant4 needs to apply a modification to avoid the nonphysical distortion. This investigation can provide an important reference for the simulation study of the polarimeter. Finally, the detection capability of 20~keV polarized photons was discussed using the modified LM model and the GD model, the modulation factor can reach 43\%.

\acknowledgments

This work was supported by the National Natural Science Foundation of China (Grant Nos. U1731239, 12027803, 11851304, U1938201), the Guangxi Science Foundation (Grant Nos. 2018GXNSFGA281007, 2017AD22006, 2018JJA110048).

\end{document}